# The History of Software Architecture - In the Eye of the Practitioner

Henry Muccini[1], Patricia Lago[2], Karthik Vaidyanathan[3],
Francesco Osborne[4], Eltjo Poort[5]

[1]University of L'Aquila, Italy
henry.muccini@univaq.it

[2]Vrije Universiteit Amsterdam, The Netherlands
p.lago@vu.nl

[3]Gran Sasso Science Institute, Italy
karthik.vaidhyanathan@gssi.it

[4]KMi, The Open University, UK
francesco.osborne@open.ac.uk

[5]CGI
eltjo.poort@cgi.com

## 1. Goals and Motivation

Software architecture (SA) is celebrating 25 years. This is so if we consider the seminal papers establishing SA as a distinct discipline[2,5], and scientific publications that have identified cornerstones of both research and practice, like architecture views, architecture description languages, and architecture evaluation[1].

With the pervasive use of cloud provisioning, the dynamic integration of multi-party distributed services, and the steep increase in the digitalization of business and society, making sound design decisions encompasses an increasingly-large and complex problem space. The role of SA is essential as never before, so much so that no organization undertakes 'serious' projects without the support of suitable architecture practices[1]. But, ***how did SA practice evolve in the past 25 years?*** *and* ***What are the challenges ahead?***

There have been various attempts to summarize the state of research and practice of SA (see Appendix A). Still, we miss the practitioners' view on the questions above.

To fill this gap, we have first extracted the top-10 topics resulting from the analysis of 5,622 scientific papers (see Figure 3). Then, we have used such topics to design an online survey filled out by 57 SA practitioners with 5 to 20+ years of experience.

## 2. The Perception of Practitioners on 25 years of Software Architecture, and the Road Ahead

In this paper, we focus on the part of the survey that investigated what the practitioners consider as the most prominent SA topics of the past 25 years, and their perception of the most impactful topics today and in the near future.

The survey included questions about: (a) the participant's background, experience, and other demographics, (b) the current organizations they work for, and the types of projects handled over the years, (c) the SA practices for the past 25 years, (d) the most impactful and trending SA topics today (with an horizon of the past 2 years), and (e) future possible directions of SA practices in industry (with a 5 years horizon). The detailed study design and questionnaire are online at http://bit.do/25yearSA. Based on the participants' answers, we have elicited the following results.

**Past**: Figure 1.PAST summarizes what the practitioners perceive as the top-10 most impactful topics of the past 25 years of SA[1]. We observe the following:

- The most-impactful topics belong to `software development process', `service oriented architecture (SOA)', `architectural styles' and `internet of things (IoT)'. They together count for 38% of the total responses.
- While `SOA' was pretty much mentioned as such, the other topics include more specific subtopics: `software development process' prominently includes Agile, Continuous Delivery & Integration, and DevOps, but also domain-driven design, the role of requirements, legacy-, risk- and quality management, and social skills; `architectural styles' include the variety of styles that shaped SAs over time, from client-server and distributed architecture, to product line architectures, MVC, multi-tier architecture, etc. Finally, `IoT' was both mentioned as such and including subtopics like digitalization, the Web, Internet, Industry 4.0 and most prominently (an maybe surprising) mobile and mobile-first.

Where applicable, in Figure 1.PAST the number in RED indicates if that topic appeared in the top-10 topics of the scientific literature (as per Figure 3). Accordingly, we observe that:

- The top-4 topics in industry are also among the top-10 topics in scientific research, but with a different impact: `software development process' is #1 in industry and only #7 in research; `SOA' is #2 in both industry and research; apparently practitioners perceive `architectural styles' as sources of a greater impact (#3) than researchers do (#6).
- The real gap is represented by `architecture description & languages', which is #1 in academic research but only #8 in the eyes of practitioners. This should not come as a big surprise if we consider that notations and languages are beloved research topics that never really reached broad adoption in industry[4]. We argue, however, if this should be a reflection point on how we (researchers and practitioners together) could do better in describing and communicating SAs more effectively.

There are also topics mentioned by practitioners and not fitting in any category identified as most-impactful in academic research. The most prominent are:

- `quality of software' (#5): apparently, addressing quality as an explicit SA property is considered as a major contribution of the past 25 years. While the term `quality' is mentioned as such in most of the cases, sometimes terms such as quality compliance, performance, scalability, maintainability are used.
- `cloud' (#6) and `microservices' (#7): mostly mentioned as such, they are recognized as very impactful (with 14 responses each) already when looking at the past. As natural historical evolutions of SOA, they could be considered as belonging to the same topic - in which case they would count 42 responses and hence become the top-1 impactful topic of the past 25 years.

[1] Where applicable, we have classified the answers from practitioners in the top-10 topics emerging from the analysis of the literature (see Figure 3).

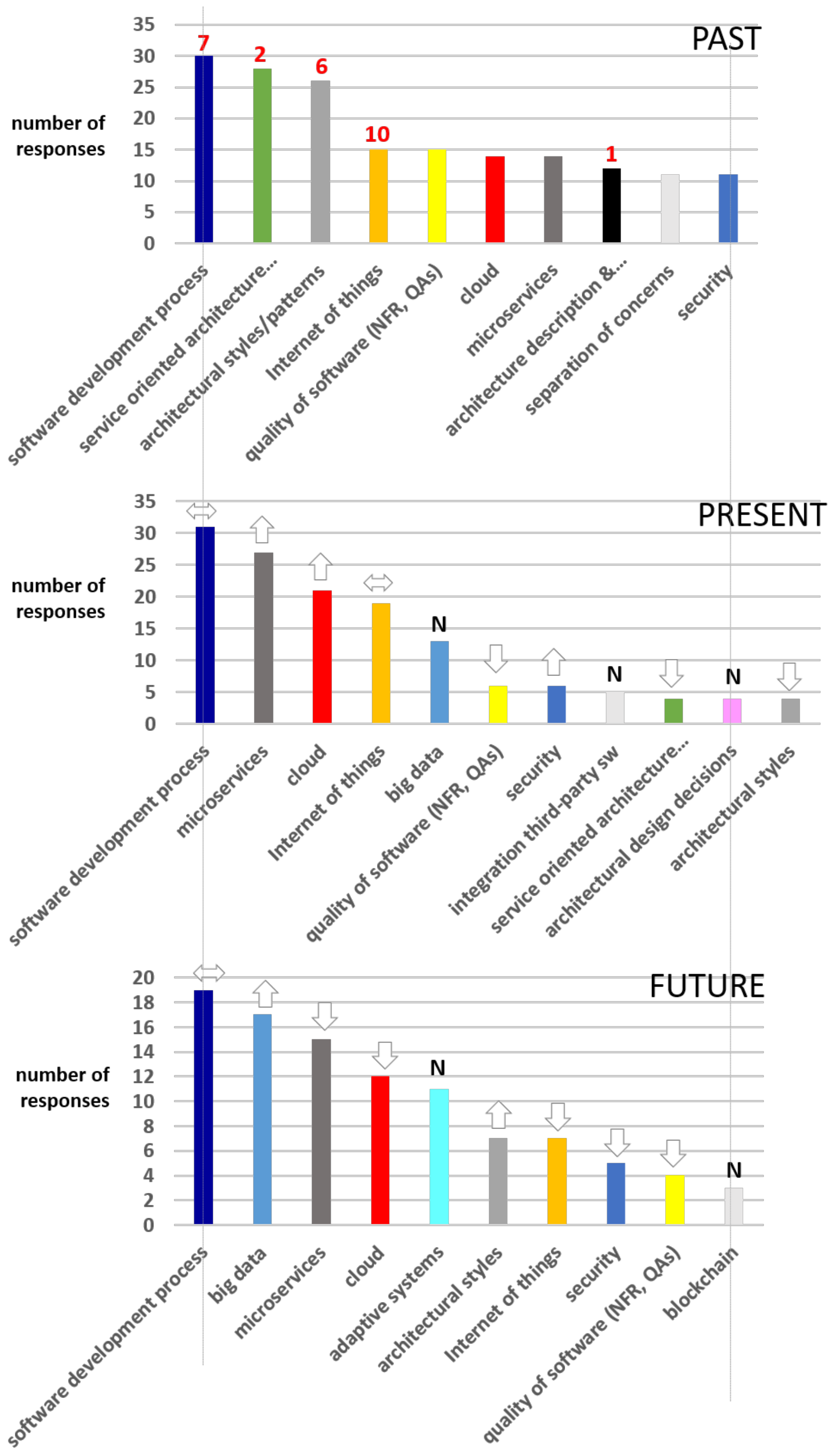

**past 25 years**
top-10 topics from practitioners.
In red, the topics position with respect to the top-10 topics from academic research.

**present**
top-10 topics from practitioners.

**future**
next 5-years' top-10 topics from practitioners. Where applicable, arrows show the change in position w.r.t. the previous period; `**N**' indicates new topics.

Figure 1: Past, present and future

**Present:** Figure 1.PRESENT runs the same kind of analysis but considers the most impactful SA topics in industry *today*. Here, `architectural styles' are relegated to position #10 (tied with `SOA' and `architectural design decisions'). On the other hand:

- `SOA' is being replaced by `cloud' (#2) and `microservices' (#3): we think this is a natural technological evolution, witnessing service orientation as a pervasive SA style across sectors and application domains.
- `software development process' remains stable (#1). With respect to the answers related to the past 25 years, today ``it's all about agility": next to agile, DevOps, and continuous-architecting, we notice an increasing awareness that architecture plays (or should play) a bigger role in steering agile development (c.f. `agile architecting', `architecture in agile environments', `emerging architecture in agile').
- Also `IoT' remains stable (#4). With respect to the past, however, it is perceived as having a higher impact, with a focus shifted from mobile devices to architecting *for* IoT. This seems to follow the trend identified by Gartner for the Top Strategic technologies for 2018[3], which includes `intelligent things' (fusing AI with IoT).
- Noticeably, both `quality of software' (#6) and `security' (#7) decrease in the perception of today's impactful topics. It would be interesting to understand if this is because architects *know* how to handle them (hence have reached maturity) or they simply perceive other topics as more important.
- Overall, today's most-impactful topics together count for about 70% of the total responses, hence showing a maybe-not-that-surprising but stronger consensus with respect to the "topics of the past" (38%). Of these, process, service orientation (i.e. cloud and microservices) and IoT together count for 62% of the total responses!

Also, noticeable *new entries* in today's top-10 topics, are:

- `big data' (#5): mentioned as such, or as AI, machine learning or analytics.
- `integration of third-party software' (#8): already mentioned in the past but at position #14 although scoring 5 responses in both cases. The specific answers, however, show a shift from inward-looking SA (e.g., EAI) to outward-looking, modern technologies (e.g., flexible contracts, adaptation, open API's).

**Future:** Naturally, when looking at the future (Figure 1.FUTURE), higher uncertainty is reflected in a larger spread of the responses over the different topics. Even if the practitioners believe that the top-4 topics will remain unchanged in the next 5 years, they account for 52% of the total responses (hence suggesting a 10% lower impact than today).

Figure 1.FUTURE shows that:

- Topics like `software development process', `big data', `microservices' and `cloud' are here to remain extremely important. However, :
    - For the `software development process', practitioners' responses show more concerns about managing an increasing complexity, possibly across organizations, and an increasing attention to higher automation.
    - For `big data', there is some mentioning of the role of AI and predictions in everyday lives.
    - `microservices' are expected to reach maturity, while new 'cloud' subtopics point towards styles/patterns for cloud-based architectures, and SA realizing XaaS business models.

- `adaptive systems' (#5), which have been a hot topic in academic research over the past few years, are expected to gain importance in industry, too.
- Among the new topics, `blockchain' is the one making the top-10 (even if with a maybe-surprising low number of responses).
- Other brand-new topics, emerging but not in the top-10, are robotics, digital transformation, intelligent connected, green software, and ethics.

## 3. Reflection and Takeaways

Next to the results presented so far, our survey also asked the practitioners to provide their view of the SA topics that in their experience, have created the major impact over the past 25 years, *broken-down in 5-year periods*.

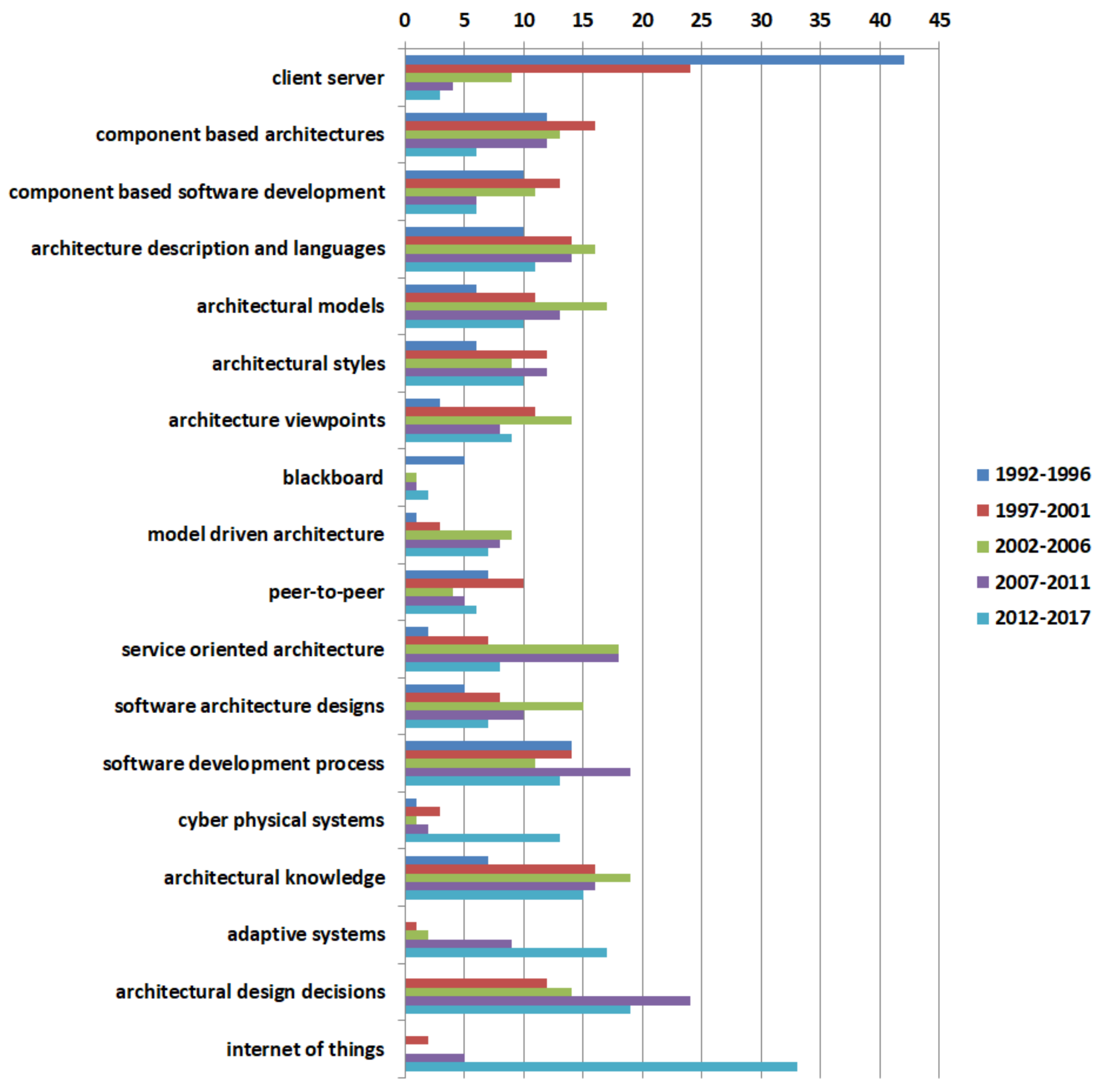

Figure 2: Mainstream industrial SA topics in 5-year periods

Figure 2 shows how topics that were mainstream in the past (e.g., client-server throughout 1992-2001) were surpassed by new topics (e.g., `architectural design decisions' and `architectural knowledge' throughout 2002-2011), and how most recent topics emerged (e.g., `cyber-physical systems' and IoT in 2012-2017). This brings us to (at least) the following takeaways:

**Takeaway 1.** In its historical journey, the notion of SA moved from the structure of a family of systems, to the holistic role of software-intensive systems in a large, complex, evolving environment. Such environment, however, is not just about technology, but encompasses people, social interaction, ecosystems of organizations, and the society at large. With a much broader focus than researchers, SA practitioners follow trends in other disciplines, from AI, IoT, and increasing adaptability, to energy and ethics. This broader focus of practitioners can be explained by a difference in drivers: whereas researchers require a well-defined scope of topics in order to get meaningful research results, in the end practitioners are held accountable for architectures that fit into the wider business and technology context of their stakeholders.

**Takeaway 2.** \`software development process' "wins it all": past, present, and future are all about how well SA steers development in increasingly agile ways, and how people skills and competencies can help. The elephant in the room is the dilemma between SA communication and its formalization: \`architecture models' and \`architecture designs' are ever-present topics that complement \`software development process' but never seem to fulfil their promise - to codify SA for dependable use, and reuse.

**Takeaway 3.** There is no *revolution* in SA topics, rather a *silent evolution* of pre-existing ones. E.g., \`software development process' *evolves* towards various forms of agility, \`architectural styles' *evolve* from \`SOA' to \`microservices' and \`cloud', \`cyber-physical systems' *evolve* into, and grow together, \`IoT\` and \`adaptive systems'. Overall, we argue a global SA trend toward flexibility, and the strenuous struggle to manage complexity.

**Takeaway 4.** Research and practice in SA are *consistently aligned*. When comparing the top-ten research topics in Figure 3[2] and the industrial mainstream topics in Figure 2, we can see that topics such as \`client server' and \`architectural styles' have trends very similar in research and practice. Topics like \`software architecture design', and \`architectural design decisions', while following different trends when comparing research and practice, are always prominent in both categories. The most visible exception is associated to \`architecture description and languages', very trendy in research, much less in practice.

[2] To extract the research topics and trends in Figure 3 we devised EDAM[6], an expert-driven automatic methodology for classifying primary studies according to their research topics. Please refer to Appendix B for more information.

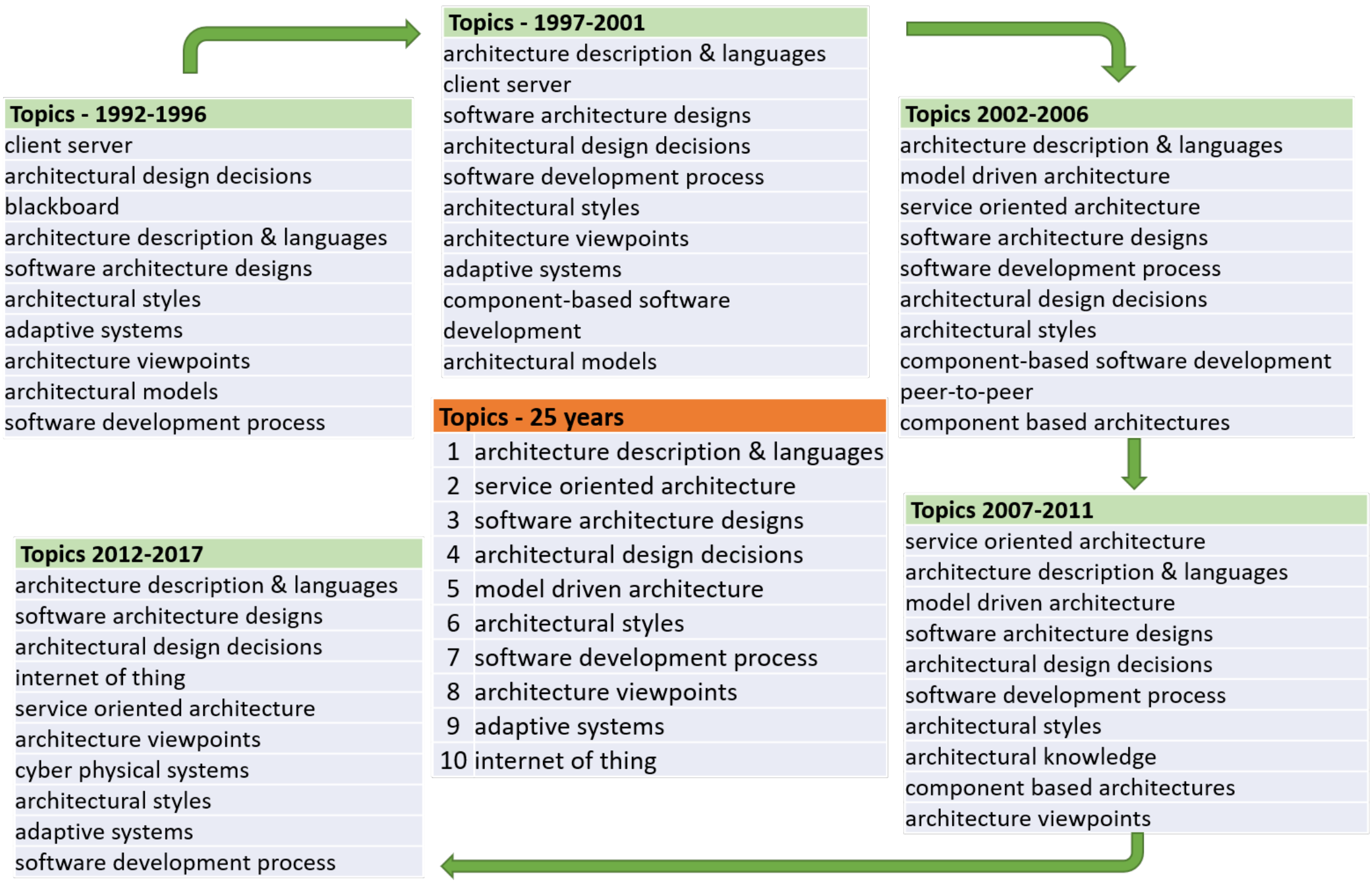

Figure 3: 25 years of SA research: top-10 topics (orange) and break-down in 5-year periods (green).

We hope this historical overview of the SA evolution brings further reflection in the readers, and inspiration.

# Appendix A. Great Papers on the History of Software Architecture

Overviews of SA research and practice have taken many different perspectives.

With the idea to provide empirical evidence (as opposed to anecdotal claim) of the benefits of SA, Galster & Weynes[3] surveyed how empirical research is applied in SA practice. With a perspective totally different from ours, they identify which types of research methods are the most popular (case studies and experiments; and involving humans as subjects or study participants). The special issue of Babar et al.[1] had a similar focus.

Various papers focused on architecture topics in practice. For instance, Woods[9] gives an helicopter view of SA from the 80's (monolithic and capturing a static structure) to this decade (intelligent and capturing fluid principles and patterns). With a specific focus on industrial practice, he reinforces topics like views and viewpoints, reuse of styles and patterns, architecture assessment techniques, and a clear communication of design decisions among stakeholders. Interestingly, he adds the increasing importance of "data and algorithms as system-wide architectural concerns" - not yet prominently addressed in research. Woods observes that the `role of the architect' is changing from a clear-cut role to a task for the many, hencewith becoming an "implicit part of mainstream practice".

A generalized shift in responsibilities emerged also from a survey of SA practitioners in 10 Countries in Europe and the US[8]. From being mainly responsible for system design, architects are now engaged in all lifecycle phases and become decision makers and knowledge managers. Topics emerging from this study include architecture as a set of design decisions, and the interplay of architecture and agility.

In their 2006 special issue, Kruchten et al.[7] provide an historical overview of 10 years of SA, with a special focus on methods and languages. The shared themes of the included papers are flexibility and adaptability, coming back in 2014[4] (with the rise of network-centric computing and cyber-physical systems) and 2015[6] (with the rise of DevOps).

Ten years later, in their editorial Hope et al.[5] reflect on the emergent need for architecture to accommodate extreme flexibility in agile development, and in embracing runtime contexts that are continuously changing and hyper-connected.

Borrmann & Paulisch[2] are among the early works extending the notion of purely-technical architecture with the challenges related to process and people: e.g., they report on the role of design decision-making for effective architecture assessment, the need for diversity in architecture reviews, and the impact of the organization structure. Among the problems, increasing complexity and need for rapid change are mentioned (topics that were meant to stay for a while, as we know by now) - and that come back in our study, too.

# Appendix B. 25 Years of Research in Software Architecture

To build a sound basis for our study, and for later comparison of the state-of-the-practice with the state-of-the-art, we extracted a list of topics characterizing the history of software architecture research. Since doing this manually is infeasible, we devised EDAM[1], an expert-driven automatic methodology for classifying primary studies according to their research topics. In short, EDAM (1) generates an ontology of research topics, (2) refines it by integrating the feedback of domain experts, and uses this knowledge-base to (3) classify papers and (4) compute relevant analytics. The reader interested in the in's and out's of EDAM can refer to the method overview[1]. The ontology and the analytics are openly available at https://tinyurl.com/y9gy2gbw. Figure B1 displays the results in terms of the top-10 topics researched in the past 25 years.

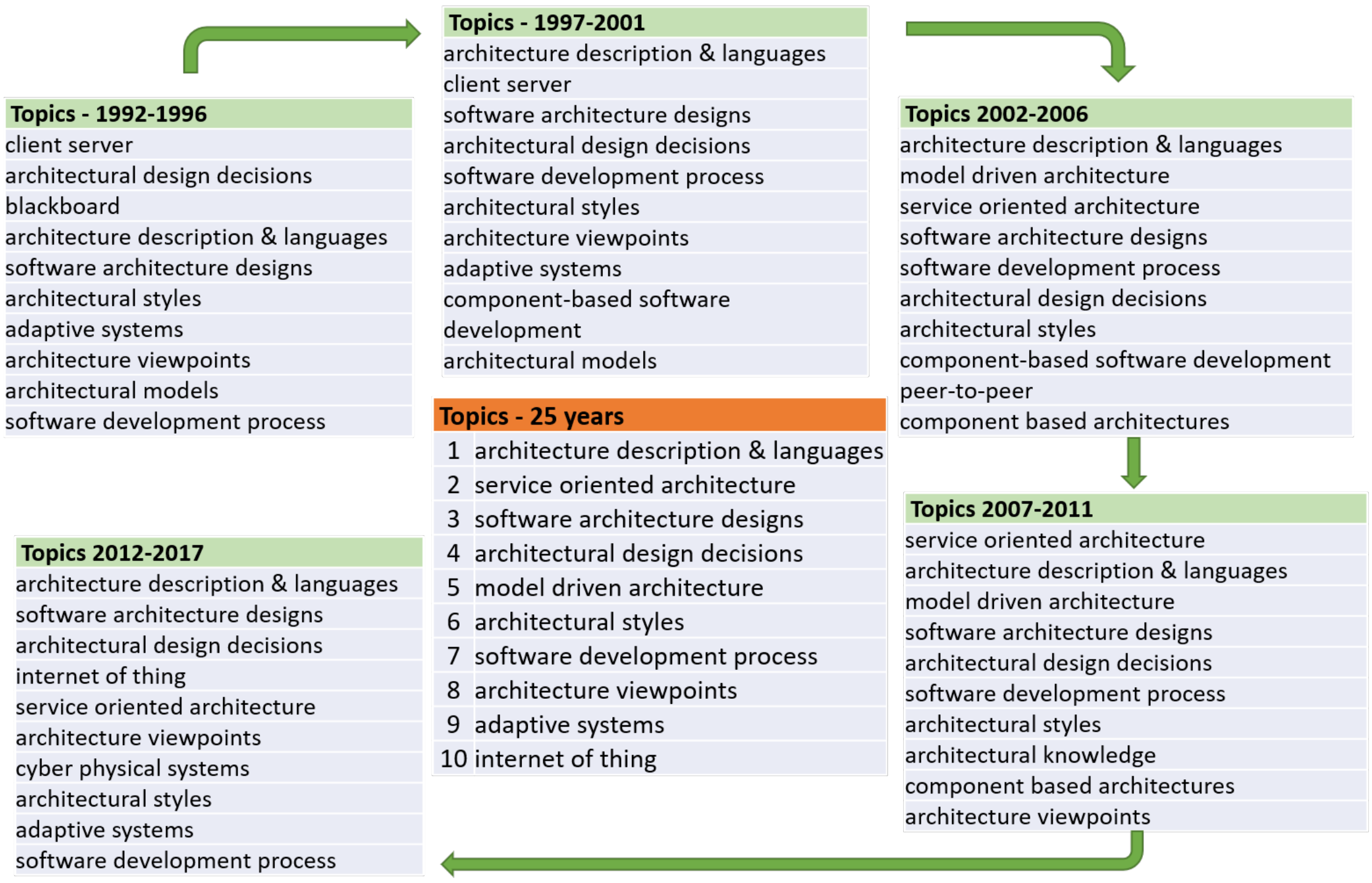

Figure B1: 25 years of SA research: top-10 topics (orange) and break-down in 5-year periods (green).

As domain ontology, we adopted the software architecture branch of the Computer Science Ontology (CSO). CSO is a large-scale ontology of research topics that was automatically generated by applying the Klink-2 algorithm[2] to 16 million publications in the Rexplore dataset and it is currently used by Springer Nature to classify proceedings in the field of Computer Science[3], such as the well-known LNCS series. We further refined the relevant branch by having it reviewed by seven senior researchers in the SA field. The resulting knowledge-base includes 86 topics on four abstraction levels.

We utilized two datasets for the analysis: a dump of the Scopus dataset in the 1997-2012 interval (SC) and a dump of the Springer Nature dataset in the 2002-2017 interval (SN). We classified the papers in both datasets by adopting a slight variation of the procedure used by the Springer Nature semantic pipeline for annotating proceedings[3]. In brief, we first extracted all papers containing in the title, abstract, or keywords one of the labels of software architecture or one of its direct subtopics (e.g., “architectural styles”, “architecture assessment”). We then associated to each topic (e.g., “architectural styles”) all papers that contain one of its labels (e.g., “architectural style”, “architectural patterns”).

We produced the topic lists depicted in Figure B1, by ranking the topics according to their number of relevant publications in a given time interval. SC and SN proved to be basically equivalent in regard to this task, yielding 95% of the same topics in the overlapping period.  We adopted SC for the four five-year intervals in the 1992-2011 period and SN for the last interval (2012-2017). The main topics of the full 25 years period were generated by summing the total values of the two databases after normalisation over their maximum number of publications per year.